\documentclass[journal,onecolumn]{rmaa}

\newcommand{\etal}{et~al.}

\newcommand{\FigBox}[2][\columnwidth]{\framebox[#1]{\rule{0pt}{#2}}}

\newcommand{\Ha}{H$\alpha$}

\title{Seyfert Galaxies with Circumnuclear/Nuclear Starbursts}

\author{Qiusheng Gu\altaffilmark{1,2},
        Deborah Dultzin-Hacyan\altaffilmark{2}
  and Jose Antonio de Diego\altaffilmark{2}}

\altaffiltext{1}{Department of Astronomy, Nanjing University, 
                 Nanjing 210093, P.R. China}
\altaffiltext{2}{Instituto de Astronom\'{\i}a, UNAM, Mexico}

\fulladdresses{
\item Qiusheng Gu :  
Current add.: Instituto de Astronom\'{\i}a,
	    Universidad Nacional Aut\'onoma de M\'exico,
	    Apdo Postal 70-264, Mexico D.F. 04510,
	    M\'exico (qsgu@astroscu.unam.mx);\\
Permanent add.: Department of Astronomy, Nanjing University,
            Nanjing 210093, P.R. China (qsgu@nju.edu.cn).
\item Deborah Dultzin-Hacyan and Jose Antonio de Diego : 
            Instituto de Astronom\'{\i}a,
            Universidad Nacional Aut\'onoma de M\'exico, 
            Apdo Postal 70-264, Mexico D.F. 04510,
            M\'exico (deborah,jdo@astroscu.unam.mx).
  }

\shortauthor{Q. Gu \etal}
\shorttitle{Seyfert Galaxies with Nuclear Starbursts}

\ReceivedDate{------} 
\AcceptedDate{------} 

\resumen{
En este art\'{\i}culo, presentamos nuestros resultados preliminares de
galaxias Seyfert con \emph{starburst} circumnuclear/nuclear. Hemos
buscado la literatura reciente y hemos encontrado 76 galaxias
activas con evidencia de actividad starburst nuclear, de las
cuales 16 son Seyfert 1, 51 Seyfert 2, y 9 LINERs. Despu\'es de
estudiar las 51 Seyfert 2, encontramos que aquellas Seyfert 2 con
n\'ucleo Seyfert 1 oculto, tienen propiedades infrarrojo-radio
similares a las galaxias Seyfert 1, y son diferentes de las
Seyfert 2 "verdaderas" sin n\'ucleo Seyfert 1 oculto. Estas \'ultimas
son similares a las galaxias starburst. }

\abstract{
  In this paper, we present our preliminary results on Seyfert galaxies 
  with circumnuclear/nuclear starburst(SB) activity. We have searched 
  the recent available literature and found 76 active galaxies with 
  clear evidence of nuclear SB activity, among which 16 are
  Seyfert 1s, 51 Seyfert 2s, and 9 LINERs. After studying the 51 Seyfert 
  2s, we find that those Seyfert 2s with hidden Seyfert 1 nuclei, have 
  similar Infrared-Radio properties as Seyfert 1 galaxies, and are 
  different from "real" Seyfert 2s without a hidden Seyfert 1 nucleus. 
  The later are similar to starburst galaxies. 
  }
      

\begin{document}

\maketitle

\section{Introduction}

The connection between starburst(SB) activity and AGNs is one of the most 
important and hotly debated issues in the study of active galaxies. 
On the theoretical side, Norman \& Scoville (1988) have suggested that
the circumnuclear star-forming activity could drive gas into the innermost 
nuclear region to feed the central black hole. 
However, Terlevich (1992) and Terlevich \etal \ (1992)
have proposed that Seyfert 2s and LINERs could be well reproduced by
violent star formation in a metal-rich environment, such as in
the nuclei of early-type galaxies, which has been well known as the
Starburst model for AGN (see e.g. the recent review by Cid Fernandes  
1997 and references therein).  In a series of papers(Dultzin-Hacyan  1995,
Dultzin-Hacyan \& Ruano  1996, and Dultzin-Hacyan \etal \  1999), it has been 
shown that there are several lines of evidence against the simplest
formulation of a "Unified Scheme" for Seyfert galaxies, according to
which, {\it all} Seyfert 2s have a hidden Sy 1 nucleus. In this paper,
we add new evidence to the existence of two kinds of Sy 2s: The hidden
Sy 1, and the "real" or "pure" Sy 2. According to these authors the 
difference between Sy 1 and "real" Sy 2 is relative decrease in the 
accretion power in the first versus a relative increase of 
nuclear/circumnuclear SB activity in the second.

On the other hand,
more and more high-quality observations from HST and large telescopes on
ground, support the presence of active star formation activity around 
the nuclei in dozens of Seyfert galaxies. For example, in NGC 1068,
a prototype of Seyfert 2 galaxy, the circumnuclear starburst accounts for
at least 50\% of the whole bolometric luminosity, 81\% at far UV($\sim$ 1500
\AA), and 83 \% at near UV ($\sim$ 2500 \AA) bands(Gonzalez Delgado  \etal \ 
1997). And Heckman \etal \ (1997) and Gonzalez Delgado \etal \ (1998) have 
presented high resolution UV images of 4 Seyfert 2 galaxies, and found 
compact nuclear starburst in all of them.  In Seyfert 2 galaxies with 
nuclear starburst, the observed UV fluxes, as suggested by 
Colina \etal \ (1997), are dominated not by AGN, but by the young massive 
stars in the nuclear region.

There are several ways to detect star-forming activities in Seyfert
galaxies (Wilson  1987). Narrow-band \Ha \ imaging (Evans \etal \  1996,
Gonzalez Delgado \etal \  1997), high-resolution radio continuum 
mapping(Forbes \& Norris  1998), high S/N spectroscopy on central
region and using stellar population synthesis technique (Schmitt 
\etal \  1999, Boisson \etal \  2000, and Jimenez-Benito \etal \  2000),
the shape of the IRAS spectra, and infrared(IR) spectroscopics searching
for absorption lines from red supergiants (Terlevich \etal \ 1990, 
Oliva \etal \ 1995).

In order to study the role of starburst in Seyfert galaxies and the
connection between starburst and AGNs, we collect all active galaxies 
with circumnuclear and/or nuclear starburst as completely as possible from
the recent literature, which are listed in Table 1.

This paper is organized as follows. In Section 2, we present the 
basic information of our sample of 76 active galaxies with 
circumnuclear/nuclear starburst, and the IR and radio properties of 
51 Seyfert 2 galaxies 
are presented and discussed in Section 3. Finally, Section 4 
summarizes our main results.

\section{The Sample of Active Galaxies}

From the recent literature, we collect the active galaxies with 
the clear evidence of nuclear and/or circumnuclear starburst 
activities, such as nuclear star-forming rings, spectroscopic
signature (high order Balmer absorption lines, Wolf-Rayet 
features) of young massive stars, etc, and find 76 such active 
galaxies, which are presented in Table 1.

Table 1 contains the list of all 76 active galaxies with 
increasing right ascension, which includes: galaxy name (column 1); 
2000 right ascension and declination (columns 2 and 3); redshift (4) 
and activity type (5) from NASA/IPAC Extragalactic Database(NED); 
distance (6), for nearby galaxies (z $<$ 0.01), it is taken from 
the database of the CfA redshift survey, and for those with z $>$ 
0.01, it is calculated by $ V_{\rm GSR}/h_{0}$, where h$_{0}$ is 
equal to 75 km/s/Mpc; 
B magnitude (7) and the recessional velocities corrected to the 
Galactic Center (V$_{\rm GSR}$)(8) from RC3 (de Vaucouleurs \etal \ 1991),
and finally references to evidence of starburst in column 9.

Among these 76 active galaxies, the activity types of NGC 2681 and NGC 6574 
in NED are given to be "Sy", so we assign NGC 2681 to be LINER as in
Veron \& Veron (2000), and NGC 6574, "Sy2", as in Kotilainen \etal \ (2000).
Following Maiolino \& Rieke (1995), we take Seyfert 1 + 1.2 + 1.5 as
Seyfert 1 galaxies, and Seyfert 1.8 + 1.9 + 2.0 as Seyfert 2s, so we 
get 16 Seyfert 1s, 51 Seyfert 2s,
and 9 LINERs. Though our sample cannot be considered complete in any way, 
and there exists an observational bias (it is difficult to detect the 
nuclear starburst in Seyfert 1s), our results confirm that nuclear 
starburst activities are more common in Seyfert 2s than in Seyfert 1s 
(Gonzalez Delgado  \etal \  1997, Maiolino \etal \  1997, and
 Dultzin-Hacyan \etal \  1999). 

In Table 2, we summarize the IRAS fluxes and 1.49 GHz radio emission 
from the NRAO/VLA Sky Survey (NVSS)(Condon \etal \  1998), and 
T, the numerical index of Hubble type, incl, the inclination angle, 
and W20, 21-cm HI line width at 20 percent of the peak,
are taken from the Lyon-Meudon Extragalactic Database (LEDA). 

In the following text, we define "SS" for Seyfert 2 galaxies with 
Starbursting activity and "PA" for Seyfert 2s with Pure AGN activity
in order to avoid confusion.

\section{Results}

\subsection{Two types of Seyfert 2s ?}

The discovery of broad polarized Balmer lines in some Seyfert 2 galaxies 
(Antonucci \& Miller 1985, Miller \& Goodrich 1990, and Tran 1995a,b,c), 
leads to the now well-known standard unified model for Seyfert 
galaxies, where Seyfert 1 and 2 galaxies are intrinsically the same,
the observed differences are just the result of varying orientation
relative to the line of sight, and the broad-line region in Seyfert 2s
is hidden by an obscuring dusty molecular torus whose size is around
a few pc. If it were the case, there would be no statistical difference
between their host galaxies. However, recent studies have
indicated that the host galaxies of Seyfert 1 are different from those of 
Seyfert 2s, including (1) the Hubble type, Seyfert 1's is earlier than Seyfert
2's ( Malkan \etal \ 1998, Hunt \& Malkan 1999); (2) the environment,
the relative densities of Seyfert 2s are 1.6 to 2.7 times higher than those
of Seyfert 1s (Laurikainen \& Salo 1995, Dultzin-Hacyan \etal \ 1999);
(3) the circumnuclear star-forming activity, there is more 
starburst activity in Seyfert 2s than in Seyfert 1s
(Maiolino \etal \ 1997, Ohsuga \& Umemura 1999); and (4) the bar 
percentage in Seyfert 2s is also higher than that in Seyfert 1s
(Pogge 1989, Maiolino \etal \ 1997). All these evidence can't be 
explained by this simplified unified model.

On the other hand, Hutchings \& Neff (1991) and Neff \& Hutchings(1992) 
even earlier have 
suggested that there are two distinct types of Seyfert 2s, those with 
hidden Seyfert 1 nuclei and those without. 
During studying the origin of the ultraviolet(UV)continuum in Seyfert 2 
galaxies, Heckman \etal \ (1995) have also suggested the possibility of two 
entirely different kinds of type 2 Seyfert galaxies(see also Antonucci 1993).

Among these 51 Sy 2s, there are 9 PA galaxies as indicated 
 in Veron \& Veron (2000) and 42 SS galaxies, where 8 sources with upper 
 limits of 25 $\mu$m fluxes, so we need to use the survival analysis 
 methods (ASURV Rev 1.2, Isobe, Feigelson \& Nelson 1986) for the
 following statistical study. We find that the probability for these two 
 samples (PA and SS) to be extracted from the same parent population
 is less than 0.0001 \%, and the mean values of $s_{25\mu m}/s_{60\mu m}$
 are 0.484$\pm$0.082 and 0.170$\pm$0.018 for PA and SS, respectively.

Following Hutchings \& Neff (1991), we plot these 51 Seyfert 2s in Fig 1, 
where we show the distribution of 60 $\mu$m luminosity vs. the flux ratio of 
25 $\mu$m to 60 $\mu$m. 
As predicted, most of them are located in the SS region. 
We find that there are 12 sources located in the PA region, eight of them 
(NGC 262, NGC 1068, NGC 1275, NGC 4507, NGC 5506, 
IC 3639, Mark 477 and Mark 1210) have been detected polarized broad 
emission lines, except IC 4995, Mark 607, NGC 5347 and NGC 7672. 
It is very interesting to notice that the 60 $\mu$m luminosities of 
these four sources are all less than 6.3 $\times 10^{9}$ L$_\odot$, 
and also less than that of those with detectable hidden BLRs, which 
indicate that there might also exist hidden BLRs, but too faint to 
be detected (David Alexander, 2000, private communication). 
It is consistent for NGC 7672 that 
according to Miller \& Goodrich (1990), NGC 7672 might have hidden BLRs, 
but at a flux level below the detection threshold. 
For NGC 5347, Gonzalez Delgado \& Perez (1996) have observed 
very-high-excitation lines, such as [Ne V], He II, [Fe VII] and [Fe X], 
which are photoionized by a hard AGN-like continuum. 
For the other two sources, 
IC 4995 and Mark 607, there are no spectropolarimetric observations yet. 
Thus, our results confirm that there might exist two distinct Seyfert 
2s: one is the "pure" Seyfert 2 and the other is the hidden Seyfert 1.

\begin{figure}
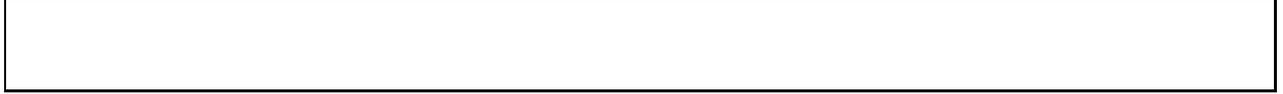

\FigBox{1cm}
\caption{Distribution of Seyfert 2 galaxies with nuclear
 starburst in the IR diagram. Open circle represents Seyfert 2 galaxy 
 without hidden Seyfert 1 nucleus, and filled circle for Seyfert 2 with 
 hidden Seyfert 1 nucleus.}
\end{figure}

\subsection{IR and Radio properties}

In order to find out the differences between PA and SS galaxies, we 
study their radio and IR properties. 
It is well known that there exists a tight relation between radio and far
infrared (FIR) emission for normal, starburst and Seyfert galaxies
(the later with more scatter). We define q to be the ratio of FIR to 1.49 GHz
emission as in Helou \etal \ (1985), which is :

\begin{equation}
q = \log \frac{\rm FIR/3.75 \times 10^{12} Hz}{S_{\nu}(1.4GHz)}
\end{equation}

\noindent Where 
$\rm FIR = 1.26 \times 10^{-14} (2.58 \times s_{60\mu m} + s_{100 \mu m})$,
s$_{60\mu m}$ and s$_{100 \mu m}$ are IRAS fluxes at 60 $\mu m$ and 100
$ \mu m$, respectively. 
Since IRAS was taken with a fixed aperture (at 60 $\mu$m, its resolution
is 1.5' $\times$ 4.75', Neugebauer et al. 1984), in general, for the most 
nearby galaxies, the circumnuclear starburst rings could lie outside the 
aperture. 
However, introspection of our sample shows that for the three
closest galaxies in our sample: Circinus (4.2 Mpc), 
NGC 4945 (3.7 Mpc), and NGC 5128 (4.9 Mpc), their star-forming rings are
located at about 10, 5.6 and 50 arcseconds, respectively (Marconi et al. 1994, 
Marconi et al. 2000a,b), which are far smaller than the IRAS aperture. And for 
the rest of galaxies, the distance is more than 10 Mpc, and 1.5' corresponds 
to larger than 4.36 Kpc. So the IRAS aperture is not a problem for our 
sample of galaxies.

For (radio-loud) AGN, q is less than 2.0 (Sopp \& Alexander 1991), 
while in starburst it is around 2.21 (Condon \etal \ 1991, 
Forbes \& Norris \ 1998)
\footnote{Sopp \& Alexander (1991) have shown that in the case
of radio-loud AGN, q is less than 2 while Condon et al. (1991) have 
suggested that "monsters" span a wide range in q values. Monsters in 
radio-selected samples generally have q $<$ 2 and can be easily recognized.
Most recently, Ji \etal \ (2000) have used q to study LINERs, and found 
that the AGN- and starburst-supported LINERs can be distinguished by their 
FIR-to-radio ratio.}.
On the other hand, according to de Grijp \etal \ (1985), 
the spectral index between 25 $\mu$m and 60 $\mu$m, $\alpha$(25,60) defined
by $f_{\nu} \propto  \nu^{\alpha} $, is also a powerful parameter for 
discriminating AGNs from starburst, with AGN-like colour, 
-1.5 $<$ $\alpha$(25,60) $<$ 0.0. Thus, we plot 51 Seyfert 2s in the
diagram q vs. $\alpha$(25,60) (see Figure 2). 

In Fig. 2, we define two regions: 
the SB region ($\alpha$(25,60) $<$ -1.5 and 2.0 $<$ q $<$ 2.8 ) and
the AGNs region ( $\alpha$(25,60) $>$ -1.5 and q $<$ 2.0 ). 
It is consistent with Fig. 1 in that most of SS galaxies are located in the 
SB region, since in these galaxies, both the FIR and radio emissions are 
dominated by star-forming activities (Sopp \& Alexander 1991, Roy et al. 1998).

There are eight galaxies in the upper-left AGNs region, seven of them
 are Seyfert 2s with hidden Seyfert 1 nuclei (NGC 262, NGC 1068,
 NGC 1275, NGC 4507, NGC 5506, Mark 477 and Mark 1210). 
 The only one without detection of hidden BLRs is NGC 3393, which 
 might be due to the huge amount of obscuring matter in the line of sight
 ($\rm N_{\rm H} > 10^{25} cm^{-2}$, Bassani et al. 1999), and recently, 
 Cooke \etal \ (1997) report that they have detected the broad emission 
 lines, so it might contain a Seyfert 1 nucleus in the center. 
 There is also one exception with the obscured Seyfert 1 nucleus but lying 
 outside the AGNs region, which is IC 3639. But it is located very nearby. 
 For four galaxies with
 possible weak hidden BLRs as indicated in Fig. 1, there is no 1.49GHz
 observation data for IC 4995, NGC 7672 is much closer to the AGNs region, 
 and the other two sources (NGC 5347 and Mark 607), due to the weakness of 
 the hidden Seyfert 1 nuclei and the presence of nuclear starburst, they 
 locate neither in the AGNs region nor the starburst region. 
 There is one galaxy (NGC 4639) to the right side of SB region, which has
 the largest FIR excess, and needs further consideration.
 It is shown that these galaxies have smaller q, which means that they
 have a radio excess, which might be related to the central AGN (Roy et
 al. 1998).

It indicates again that SS galaxies have similar properties as 
starburst galaxies, while PA galaxies are similar to Seyfert 1 galaxies.

\begin{figure}
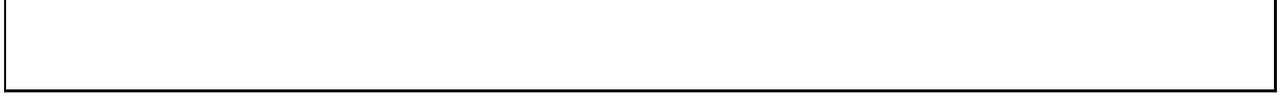

\FigBox{1cm}
\caption{Distribution of Seyfert 2 galaxies with nuclear
  starburst in the diagram of ratio of FIR to radio (q) vs. the IR 
  spectral index. The symbols have the same meaning as in Fig. 1.}
\end{figure}

\section{conclusions}
We have collected 76 active galaxies with circumnuclear or nuclear
star-forming activities, among which, 16 are Seyfert 1s, 51 Seyfert 2s
and 9 LINERs. After studying their IR and radio properties 
of 51 Seyfert 2s, we confirm that there are two distinct types
of Seyfert 2 galaxies, those with hidden Seyfert 1 nuclei share 
similar properties with Seyfert 1s, and those without, are similar to
Starburst galaxies.

\acknowledgments

 The authors are very grateful to the anonymous referee and Dr. Jorge 
 Cant\'{o} Illa for their
 careful reading the manuscript and valuable comments, which 
 improved both the contents and the presentation.
 We also would like to thank David Alexander for helpful discussion. This 
 work is supported by grant IN 115599 from PAPIIT-UNAM and a grant from
 the NSF of China, QSGU acknowledges support from UNAM postdoctoral program 
 (Mexico) and from the National Major Project for Basic Research of the
 State Scientific Commission of China. This research 
 has made use of NASA's Astrophysics Data System Abstract Service and
 the NASA/IPAC Extragalactic Database (NED) which is operated by the Jet
 Propulsion Laboratory, California Institute of Technology, under contract 
 with the National Aeronautics and Space Administration. We have also
 made use of the LEDA database, www-obs.univ-lyon1.fr.

\begin{table}
\begin{center}
\caption{Active Galaxies with Circumnuclear/Nuclear Starbursts}
\begin{tabular}{lccclrrrl}
\hline
\hline
Galaxy Name & $\alpha$(2000) & $\delta$ (2000) & z & Type& Dist$^{a}$ & B$_{\rm T}^{0}$ & V$_{\rm GSR}$ & Ref$^{b}$ \\ \hline
Mark 334 & 00:03:09.6 & +21:57:36.6 &0.02196 & Sy1.8 &90.2 &        &    6763  & GD97 \\
NGC  262 & 00:48:47.1 & +31:57:25.1 &0.01503 & Sy2   & 62.3 &  13.94 &    4669  & GD97 \\
IC  1816 & 02:31:50.9 & $-$36:40:16.2 &0.01695 & Sy1   & 67.8 &  13.66 &    5086  & Sc99\\
NGC  985 & 02:34:37.8 & $-$08:47:15.4 &0.04314 & Sy1   & 172.2&  13.89 &   12916  & Ar99\\
NGC 1068 & 02:42:40.7 & $-$00:00:47.8 &0.00379 & Sy2   & 14.4 &   9.47 &    1144  & GD97 \\
NGC 1097 & 02:46:19.1 & $-$30:16:28.0 &0.00425 & Sy1   & 14.5 &   9.92 &    1255  & St96a\\
NGC 1144 & 02:55:12.2 & $-$00:11:00.8 &0.02885 & Sy2   & 115.3&  13.21 &    8646  & Gao97\\
Mark 1066 & 02:59:58.6 & +36:49:14.3 &0.01202 & Sy2  & 49.4 &  13.04 &    3705  & GD00\\
NGC 1275 & 03:19:48.2 & +41:30:42.1 &0.01756 & Sy2   & 71.5 &        &    5362  & Pr87\\
NGC 1326 & 03:23:56.4 & $-$36:27:51.6 &0.00454 & LINER & 16.9 &  11.39 &    1244  & St96b\\
Mark  607 & 03:24:48.7 & $-$03:02:32.7 &0.00906 & Sy2  & 35.7 &        &    2674  & Sc99\\
NGC 1365 & 03:33:36.4 & $-$36:08:25.5 &0.00546 & Sy1.8 & 16.9 &   9.93 &    1541  & Fo98\\
NGC 1386 & 03:36:45.4 & $-$35:59:57.0 &0.00290 & Sy2   & 16.9 &  12.12 &     741  & Sc99\\
NGC 1598 & 04:28:33.4 & $-$47:46:56.1 &0.01712 & LINER & 65.9 &  13.44 &    4939  & St96b\\
NGC 1672 & 04:45:42.1 & $-$59:14:56.9 &0.00450 & Sy2   & 14.5 &  10.25 &    1155  & de99\\
NGC 1667 & 04:48:37.1 & $-$06:19:11.9 &0.01517 & Sy2   & 59.5 &  12.41 &    4459  & Fo98\\
NGC 1808 & 05:07:42.3 & $-$37:30:45.9 &0.00334 & Sy2   & 10.8 &  10.45 &     835  & Ko96\\
ESO 362-G08&05:11:09.0 & $-$34:23:35.9 &0.01596 & Sy2  & 61.5 &  13.51 &    4616  & Sc99\\
ESO 362-G18&05:19:35.8 & $-$32:39:30.9 &0.01264 & Sy1.5& 48.0 &  13.58 &    3603  & Ts95\\
Mark 1210 & 08:04:05.8 & +05:06:49.7 &0.01350 & Sy2  & 52.1 &  14.21 &    3910  & St99\\
NGC 2639 & 08:43:38.1 & +50:12:20.0 &0.01113 & Sy1.9 & 44.2 &  12.19 &    3315  & GD97\\
NGC 2681 & 08:53:32.8 & +51:18:50.0 &0.00231 & LINER$^{c}$&13.3&10.88&     725  & GD97\\
NGC 2782 & 09:14:05.1 & +40:06:49.4 &0.00855 & Sy1   & 37.3 &  12.01 &    2551  & Yo99\\
NGC 3081 & 09:59:29.5 & $-$22:49:34.6 &0.00796 & Sy2   & 32.5 &  12.59 &    2164  & St96b\\
NGC 3147 & 10:16:53.6 & +73:24:02.7 &0.00941 & Sy2   & 40.9 &  11.24 &    2935  & La97\\
NGC 3185 & 10:17:38.5 & +21:41:18.1 &0.00406 & Sy2   & 21.3 &  12.65 &    1159  & GD97\\
NGC 3227 & 10:23:30.6 & +19:51:54.0 &0.00386 & Sy1.5 & 20.6 &  11.18 &    1079  & GD97\\
NGC 3393 & 10:48:23.4 & $-$25:09:42.8 &0.01251 & Sy2   & 47.1 &  12.64 &    3532  & Ts95\\
NGC 3660 & 11:23:32.2 & $-$08:39:30.2 &0.01227 & Sy2   & 47.1 &        &    3529  & GD97\\
Mark  423 & 11:26:48.5 & +35:15:03.1 &0.03227 & Sy1.9& 128.7&  14.64 &    9650  & Ra93\\
NGC 3758 & 11:36:29.0 & +21:35:47.8 &0.02985 & Sy1   & 118.2&  15.00 &    8865  & Ra93\\
NGC 3783 & 11:39:01.8 & $-$37:44:18.7 &0.00973 & Sy1   & 38.5 &  12.04 &    2728  & Ts95\\
NGC 3982 & 11:56:27.9 & +55:07:36.2 &0.00370 & Sy2   & 17.0 &  11.68 &    1184  & Fo98\\
NGC 4303 & 12:21:54.9 & +04:28:25.1 &0.00522 & Sy2   & 15.2 &  10.12 &    1486  & Co99\\
NGC 4314 & 12:22:31.9 & +29:53:43.3 &0.00321 & LINER &  9.7 &  11.17 &     965  & GD97\\
NGC 4321 & 12:22:54.9 & +15:49:20.8 &0.00524 & LINER & 16.8 &   9.98 &    1540  & Kn98\\
NGC 4507 & 12:35:36.5 & $-$39:54:33.3 &0.01180 & Sy2   & 44.3 &  12.28 &    3320  & Ts95\\
NGC 4593 & 12:39:39.4 & $-$05:20:39.3 &0.00900 & Sy1   & 39.5 &  11.43 &    2393  & GD97\\
IC  3639 & 12:40:52.9 & $-$36:45:21.5 &0.01092 & Sy2   & 41.8 &  12.71 &    3137  & St99\\
NGC 4639 & 12:42:52.4 & +13:15:27.2 &0.00337 & Sy1.8 & 16.8 &  11.85 &     963  & GD97\\
NGC 4736 & 12:50:53.1 & +41:07:13.6 &0.00103 & LINER &  4.3 &   8.75 &     360  & GD97\\
NGC 4945 & 13:05:27.5 & $-$49:28:05.6 &0.00187 & Sy2   &  3.7 &   7.43 &     383  & Co97\\
NGC 5128 & 13:25:27.6 & $-$43:01:08.8 &0.00183 & Sy2   &  4.9 &   7.30 &     398  & Ma00\\
NGC 5135 & 13:25:43.9 & $-$29:50:02.3 &0.01372 & Sy2   & 52.8 &  12.37 &    3959  & GD98\\
\hline
\hline
\end{tabular}
\end{center}
\end{table}

\newpage
\begin{center}
TABLE 1. {\it continued}
\begin{tabular}{lccclrrrl}
\hline
\hline
Galaxy Name & $\alpha$(2000) & $\delta$ (2000) & z & Type& Dist$^{a}$ & B$_{\rm T}^{0}$ & V$_{\rm GSR}$ & Ref$^{b}$ \\ \hline
NGC 5194 & 13:29:52.4 & +47:11:53.8 &0.00154 & LINER &  7.7 &   8.67 &     551  & GD97\\
NGC 5248 & 13:37:32.2 & +08:53:05.9 &0.00385 & Sy2   & 22.7 &  10.63 &    1128  & PR98\\
NGC 5347 & 13:53:17.8 & +33:29:27.0 &0.00779 & Sy2   & 36.7 &  13.10 &    2443  & GD98\\
NGC 5427 & 14:03:25.6 & $-$06:01:42.3 &0.00873 & Sy2   & 38.1 &  11.73 &    2592  & GD98\\
Circinus & 14:13:09.3 & $-$65:20:20.6 &0.00145 & Sy2   &  4.2 &   8.50 &     268  & Co97\\
NGC 5506 & 14:13:14.9 & $-$03:12:27.2 &0.00618 & Sy1.9 & 28.7 &  12.26 &    1782  & Ma94\\
NGC 5643 & 14:32:40.7 & $-$44:10:28.5 &0.00400 & Sy2   & 16.9 &  10.23 &    1066  & Sc99\\
Mark 477 & 14:40:38.1 & +53:30:16.0 &0.03780 & Sy2   & 153.5&  15.23 &   11511  & He97\\
NGC 5728 & 14:42:23.9 & $-$17:15:11.0 &0.00930 & Sy2   & 42.2 &  11.72 &    2735  & Ca96\\
NGC 5953 & 15:34:32.4 & +15:11:37.7 &0.00656 & Sy2   & 33.0 &        &    2040  & Co97\\
NGC 6221 & 16:52:46.7 & $-$59:12:59.0 &0.00494 & Sy2   & 19.4 &   9.77 &    1368  & Co97\\
NGC 6300 & 17:16:59.2 & $-$62:49:11.2 &0.00370 & Sy2   & 14.3 &  10.20 &     997  & Ev96\\
NGC 6574 & 18:11:51.3 & +14:58:52.1 &0.00761&Sy2$^{d}$&35.0 &  11.79 &    2441  & Ko00\\
NGC 6764 & 19:08:16.4 & +50:55:59.6 &0.00806 & Sy2   & 37.0 &  12.14 &    2637  & Bo90\\
NGC 6814 & 19:42:40.6 & $-$10:19:24.6 &0.00521 & Sy1.5 & 22.8 &  11.32 &    1676  & GD97\\
NGC 6810 & 19:43:34.2 & $-$58:39:20.6 &0.00677 & Sy2   & 25.3 &  11.49 &    1888  & Co97\\
NGC 6860 & 20:08:46.1 & $-$61:05:55.9 &0.01488 & Sy1   & 58.5 &  13.21 &    4385  & Li93\\
IC  4995 & 20:19:59.1 & $-$52:37:19.8 &0.01609 & Sy2   & 62.1 &  14.08 &    4657  & Br98\\
NGC 6951 & 20:37:14.5 & +66:06:19.7 &0.00475 & Sy2   & 24.1 &  10.71 &    1645  & Pe00\\
Mark  509 & 20:44:09.7 & $-$10:43:24.5 &0.03440 & Sy1.2& 138.4&  14.41 &   10380  & Wi88\\
NGC 7130 & 21:48:19.5 & $-$34:57:09.2 &0.01615 & Sy2   & 64.7 &  12.88 &    4850  & Co97 \\
NGC 7177 & 22:00:41.3 & +17:44:16.5 &0.00384 & LINER & 18.2 &  11.47 &    1343  & Di00\\
NGC 7214 & 22:09:07.6 & $-$27:48:35.5 &0.02313 & Sy1.2 & 91.5 &  12.83 &    6865  & Ra98\\
NGC 7213 & 22:09:16.2 & $-$47:10:00.4 &0.00598 & Sy1.5 & 22.0 &  11.13 &    1767  & St96b\\
NGC 7469 & 23:03:15.6 & +08:52:26.4 &0.01632 & Sy1.2 & 67.4 &  12.64 &    5053  & GD98\\
Mark  315 & 23:04:02.6 & +22:37:27.5 &0.03887 & Sy1.5& 160.2&  14.34 &   12016  & Wi88\\
NGC 7479 & 23:04:56.6 & +12:19:22.7 &0.00794 & Sy2   & 33.9 &  11.22 &    2544  & La99\\
NGC 7552 & 23:16:11.0 & $-$42:34:59.0 &0.00529 & LINER & 19.5 &  11.13 &    1568  & Fo94\\
NGC 7582 & 23:18:23.5 & $-$42:22:14.0 &0.00525 & Sy2   & 17.6 &  10.83 &    1551  & Co97\\
NGC 7592 & 23:18:22.5 & $-$04:24:58.5 &0.02444 & Sy2   & 99.1 &        &    7429  & Ra92\\
NGC 7590 & 23:18:54.6 & $-$42:14:21.0 &0.00532 & Sy2   & 17.3 &  11.46 &    1569  & Ts95\\
NGC 7672 & 23:27:31.4 & +12:23:06.4 &0.01338 & Sy2   & 57.0 &  14.31 &    4274  & GD97\\
\hline
\hline
\end{tabular}
\end{center}

\noindent $^{a}$: in unit of Mpc; $^{b}$: References for evidence of starburst: 
 Ar99 = Arribas  \etal \  1999; Bo90 = Boer  \etal \  1990;
 Br98 = Bransford  \etal \  1998; Ca96 = Capetti  \etal \  1996;
 Co97 = Colina  \etal \  1997; Co99 = Colina  \etal \  1999;
 de99 = de Naray  \etal \  1999; Di00 = Diaz  \etal \  2000;
 Ev96 = Evans  \etal \  1996; Fo94 = Forbes  \etal \  1994;
 Fo98 = Forbes  \etal \  1998;
 Gao97 = Gao  \etal \  1997; GD97 = Gonzalez Delgado  \etal \  1997; 
 GD98 = Gonzalez Delgado  \etal \  1998; GD00 = Gonzalez Delgado  2000;
 He97 = Heckman  \etal \  1997; Kn98 = Knapen  1998;
 Ko96 = Kotilainen  \etal \  1996; 
 Ko00 = Kotilainen  \etal \  2000; La97 = Laurent  \etal \  1997;
 La99 = Laine  \etal \  1999; Li93 = Lipari  \etal \  1993;
 Ma00 = Marconi  \etal \  2000b; Ma94 = Maiolino  \etal \  1994;
 Pe00 = Perez  \etal \  2000;
 Pr87 = Pronik \& Metik  1987; PR98 = Perez-Ramirez \& Knapen  1998;
 Ra92 = Rafanelli  \etal \  1992; Ra93 = Rafanelli  \etal \  1993;
 Ra98 = Radovich  \etal \  1998;
 Sc99 = Schmitt, Storchi Bergmann \& Cid Fernandes, 1999;
 St96a = Storchi Bergmann  \etal \  1996a; 
 St96b = Storchi Bergmann  \etal \  1996b;
 St99 = Storchi Bergmann  1999; Ts95 = Tsvetanov \& Petrosian  1995;
 Wi88 = Wilson  1988; 
 Yo99 = Yoshida  \etal \  1999.
 $^{c}$ is from Veron \& Veron (2000); $^{d}$ is from Ko00.

\begin{table}
\begin{center}
\caption{Basic Data for Active Galaxies}
\begin{tabular}{lrrrrrrrr}
\hline
\hline
Galaxy Name & T & incl & W20&s12$^{a}$&s25$^{a}$&s60$^{a}$&s100$^{a}$ &f149$^{b}$ \\ 
\hline
Mark 334 & 4.9  & 44.6 & 377.4 &  0.25L&    1.05 &    4.26 &    4.51 &    28.4\\
NGC  262 & -0.5 & 15.9 & 90.4  &  0.31 &    0.77 &    1.44 &    1.83 &   292.7\\
IC  1816 & 2.0  & 33.7 &       &  0.25L&    0.42 &    1.42 &    2.39 &    35.1\\
NGC  985 & 9.9  & 27.2 &       &  0.27L&    0.55 &    1.44 &    2.00 &    17.3\\
NGC 1068 & 3.0  & 31.7 & 299.7 & 38.30 &   86.83 &  185.80 &  240.50 &  4849.0\\
NGC 1097 & 3.3  & 48.9 & 398.9 &  1.84 &    5.80 &   45.85 &   83.79 &   250.2\\
NGC 1144 & -4.2 & 70.4 &       &  0.26 &    0.70 &    5.32 &   11.59 &   155.5\\
Mark 1066 & -1.1 & 54.7 &       &  0.50 &    2.31 &   10.45 &   13.10 &   100.9\\
NGC 1275 & -2.2 & 52.0 &       &  0.97 &    3.62 &    7.22 &    8.01 & 22829.7\\
NGC 1326 & -0.8 & 54.3 & 265.2 &  0.27 &    0.79 &    8.31 &   14.30 &    34.4\\
Mark  607 & 1.1  & 83.8 &       &  0.33 &    1.08 &    2.36 &    2.88 &     6.5\\
NGC 1365 & 3.2  & 58.2 & 398.7 &  3.21 &   11.09 &   78.15 &  141.50 &   376.8\\
NGC 1386 & -0.8 & 90.0 &       &  0.50 &    1.44 &    5.89 &    9.92 &    37.8\\
NGC 1598 & 5.0  & 56.1 &       &  0.25L&    0.21L&    1.45 &    4.23 &     $--^c$\\
NGC 1672 & 3.3  & 33.1 & 276.2 &  1.47 &    4.03 &   34.80 &   69.46 &     $--$\\
NGC 1667 & 5.0  & 39.7 & 324.2 &  0.38 &    0.67 &    5.85 &   14.67 &    77.3\\
NGC 1808 & 1.2  & 60.6 & 333.6 &  4.12 &   15.87 &   97.12 &  136.50 &   528.8\\
ESO 362-G08&-0.4&74.9  &       &  0.25L&    0.19L&    0.64 &    0.82 &     3.4\\
ESO 362-G18 & 0.1&58.7 &       &  0.25L&    0.59 &    1.49 &    2.02 &    15.0\\
Mark 1210 &      &  5.8 &       &  0.55 &    2.09 &    1.84 &    1.51 &   114.9\\
NGC 2639 & 1.0  & 43.3 &       &  0.35L&    0.39L&    2.00 &    7.07 &   115.7\\
NGC 2681 & 0.4  &  0.0 &       &  0.35 &    0.59 &    7.07 &   11.40 &    10.2\\
NGC 2782 & 1.1  & 49.8 & 190.3 &  0.51 &    1.47 &    8.47 &   13.81 &   125.5\\
NGC 3081 & -0.1 & 41.0 & 257.0 &  $ -^d$ &    $ -$ &    $ -$ &   $ -$  &     5.7\\
NGC 3147 & 4.0  & 31.2 & 396.8 &  0.52 &    0.61 &    6.66 &   24.63 &    92.5\\
NGC 3185 & 1.0  & 54.2 & 284.4 &  0.25L&    0.29L&    1.58 &    3.68 &     5.6\\
NGC 3227 & 1.5  & 49.2 & 418.7 &  0.67 &    1.74 &    7.98 &   17.46 &   100.7\\
NGC 3393 & 1.1  & 23.7 &       &  0.25L&    0.71 &    2.38 &    3.93 &    81.5\\
NGC 3660 & 3.9  & 36.8 & 291.9 &  0.26L&    0.25L&    1.95 &    4.65 &    14.8\\
Mark 423 & 3.0  & 57.6 &       &  0.28L&    0.25L&    1.37 &    2.30 &    14.2\\
NGC 3758 & 2.7  & 27.4 & 332.9 &  4.24L&    0.28 &    1.43 &    2.43 &    11.2\\
NGC 3783 & 1.4  & 28.5 &       &  0.77 &    2.43 &    3.37 &    5.12 &    44.6\\
NGC 3982 & 3.1  & 30.5 & 234.3 &  0.48 &    0.84 &    6.92 &   16.00 &    57.7\\
NGC 4303 & 4.0  & 25.4 & 170.9 &  0.49L&    0.61L&   23.52 &   61.69 &   435.8\\
NGC 4314 & 1.0  & 19.3 &       &  0.25L&    0.39 &    3.76 &    7.59 &    14.1\\
NGC 4321 & 4.0  & 36.8 & 269.5 &  0.79 &    1.31 &   18.23 &   57.63 &    87.1\\
NGC 4507 & 1.9  & 35.4 &       &  0.46 &    1.41 &    4.58 &    5.60 &    67.4\\
NGC 4593 & 3.0  & 47.1 & 366.8 &  0.34 &    0.92 &    2.81 &    6.00 &     4.8\\
IC  3639 & 4.0  & 20.3 &       &  0.65 &    2.30 &    7.20 &   11.14 &    89.0\\
NGC 4639 & 3.6  & 49.1 & 330.4 &  0.25L&    0.31L&    1.41 &    4.54 &     1.8\\
NGC 4736 & 2.5  & 30.1 & 232.4 &  2.78 &    3.48 &   56.07 &  105.20 &   269.9\\
NGC 4945 & 6.1  & 84.6 & 381.6 &  3.63 &   14.24 &  388.10 &  685.60 &     $--$\\
NGC 5128 & -2.3 & 50.4 & 555.6 & 11.14 &   14.99 &  171.50 &  337.60 &     $--$\\
NGC 5135 & 2.3  & 22.5 & 141.4 &  0.67 &    2.48 &   16.18 &   30.83 &   200.5\\
\hline
\hline
\end{tabular}
\end{center}
\end{table}

\newpage
\begin{center}
TABLE 2. {\it continued}

\begin{tabular}{lrrrrrrrr}
\hline
\hline
Galaxy Name & T & incl & W20&s12$^{a}$&s25$^{a}$&s60$^{a}$&s100$^{a}$ &f149$^{b}$ \\ 
\hline
NGC 5194 & 4.0  & 48.1 & 197.5 &  1.36 &    2.38 &   32.00 &  122.90 &   430.3\\
NGC 5248 & 4.0  & 40.9 & 281.7 &  0.96 &    1.49 &   17.70 &   44.00 &   160.3\\
NGC 5347 & 2.0  & 39.9 & 114.5 &  0.29 &    0.92 &    1.44 &    2.71 &     6.0\\
NGC 5427 & 5.0  & 38.9 &       &  0.28 &    0.62 &    4.86 &   16.46 &    86.2\\
Circinus & 3.2  & 67.4 & 303.1 & 18.80 &   68.44 &  248.70 &  315.90 &     $--$\\
NGC 5506 & 1.3  & 90.0 & 330.5 &  1.30 &    3.64 &    8.81 &    9.72 &   339.4\\
NGC 5643 & 5.0  & 28.8 & 205.4 &  0.86 &    3.35 &   18.71 &   44.19 &     $--$\\
Mark 477 &      & 40.5 &       &  0.25L&    0.54 &    1.35 &    1.85 &    60.8\\
NGC 5728 & 1.1  & 58.1 & 412.9 &  0.32L&    0.81 &    8.40 &   15.17 &    70.8\\
NGC 5953 & 0.2  & 46.6 & 261.3 &  0.56 &    1.07 &   10.42 &   20.31 &    91.7\\
NGC 6221 & 4.8  & 50.2 & 311.4 &  1.49 &    5.27 &   36.32 &   84.50 &     $--$\\
NGC 6300 & 3.1  & 54.2 & 326.3 &  0.74 &    2.15 &   14.05 &   40.25 &     $--$\\
NGC 6574 & 4.0  & 42.5 & 373.6 &  0.92 &    1.67 &   14.48 &   27.82 &   102.2\\
NGC 6764 & 3.6  & 58.1 & 294.0 &  0.38 &    1.33 &    6.48 &   11.90 &   110.9\\
NGC 6814 & 4.0  & 11.1 &  97.6 &  0.33 &    0.59 &    5.69 &   18.16 &    51.9\\
NGC 6810 & 2.1  & 86.0 &       &  1.05 &    3.48 &   18.12 &   34.50 &     $--$\\
NGC 6860 & 3.0  & 58.6 &       &  0.24L&    0.35 &    1.05 &    2.58 &     $--$\\
IC  4995 & -2.0 & 57.8 &       &  0.25L&    0.36 &    0.90 &    1.28 &     $--$\\
NGC 6951 & 4.0  & 26.1 & 328.8 &  0.45 &    1.17 &   13.49 &   37.14 &    70.4\\
Mark 509 &      & 36.3 &       &  0.34 &    0.74 &    1.42 &    1.43 &    19.2\\
NGC 7130 & 1.1  & 23.8 &       &  0.63 &    2.14 &   16.67 &   26.27 &   190.6\\
NGC 7177 & 2.5  & 51.3 & 312.0 &  $ -$ &    $ -$ &    $ -$ &    $ -$ &    27.7\\
NGC 7214 & 4.4  & 48.7 &       &  0.25L&    0.37 &    2.13 &    5.24 &    29.2\\
NGC 7213 & 0.9  & 29.6 & 443.9 &  0.63 &    0.74 &    2.56 &    8.63 &     $--$\\
NGC 7469 & 1.1  & 48.9 & 386.3 &  1.30 &    5.48 &   26.95 &   35.22 &   181.0\\
Mark 315 & -5.0 & 78.2 & 289.9 &  0.39L&    0.37 &    1.50 &    2.83 &    22.8\\
NGC 7479 & 4.4  & 40.9 & 376.3 &  0.75 &    3.32 &   12.12 &   24.93 &   101.4\\
NGC 7552 & 2.4  & 28.2 & 230.6 &  2.98 &   11.96 &   72.93 &  100.90 &     $--$\\
NGC 7582 & 2.0  & 68.6 & 382.4 &  1.35 &    6.33 &   48.01 &   72.76 &     $--$\\
NGC 7592 & -0.9 & 40.1 &       &  0.36L&    1.10 &    8.07 &   10.71 &    76.1\\
NGC 7590 & 4.0  & 71.3 & 458.2 &  0.52 &    0.84 &    7.39 &   18.02 &     $--$\\
NGC 7672 & 3.0  & 40.6 &       &  0.25L&    0.28L&    0.54 &    1.15 &     6.5\\
\hline
\hline
\end{tabular}
\end{center}

\noindent 
$^{a}$ in unit of Jy; 
$^{b}$ in unit of mJy; 
$^{c}$ without information of 1.49 GHz radio emission;
$^{d}$ nondetection by IRAS.

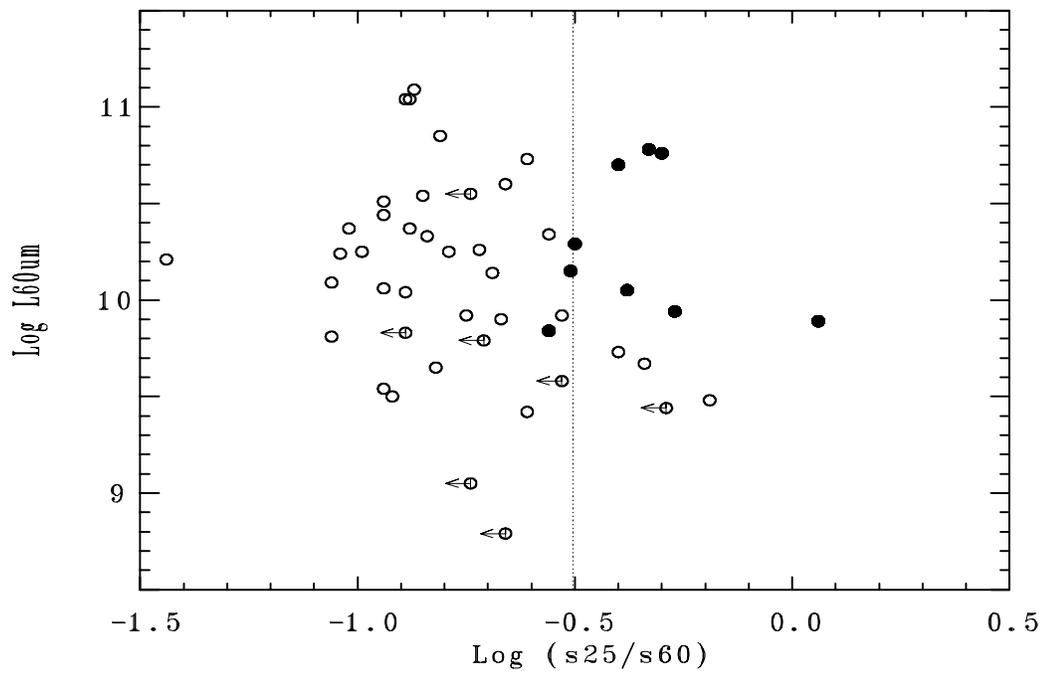
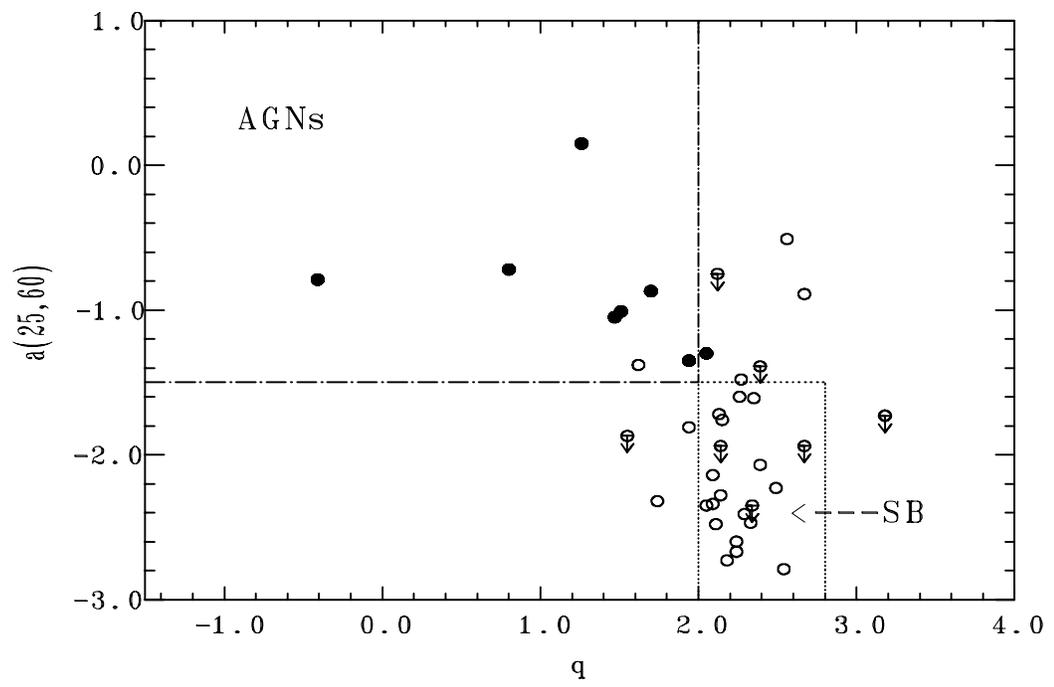
\end{document}